\documentclass{JHEP3}

\usepackage{epsfig}

\newcommand{\be}{\begin{equation}}
\newcommand{\ee}{\end{equation}}

\title{A large N phase transition in the continuum 
two dimensional SU(N) X SU(N)
principal chiral model.}
\author{R. Narayanan
\\Department of Physics, Florida International University, Miami,
FL 33199, USA\\E-mail: \email{rajamani.narayanan@fiu.edu}}
\author{ H. Neuberger
\\ Rutgers University, Department of Physics and Astronomy,
Piscataway, NJ 08855, USA\\E-mail: \email
{neuberg@physics.rutgers.edu} }
\author{ E. Vicari\\
Dipartimento di Fisica, Universit\'a di Pisa, and INFN, I-56127 Pisa, Italy\\E-mail: \email{vicari@df.unipi.it}}

\abstract {It is established by numerical means that the continuum
large N principal chiral model in two dimensions has a phase transition
in a smoothed two point function at a critical distance of the order
of the correlation length. 
}

\keywords{Large N, Lattice Gauge Field Theories}

\preprint{}

\begin{document}

\section{Introduction.}

It is well known that the principal chiral $SU(N)\times SU(N)$ model (PCM) in
two Euclidean dimensions is similar to four dimensional pure $SU(N)$ gauge
theory in many respects~\cite{review}. 

Recent numerical work provides evidence
that Wilson loops in $SU(N)$ gauge theory in two, three and four dimensions
exhibit an infinite $N$ phase transition as they are dilated from a 
small size to a large one; in the course of this dilation the eigenvalue distribution of the untraced Wilson loop unitary matrix expands from a small 
arc on the unit circle to encompassing the entire unit circle~\cite{ourjhep, three-d}.  There is
further evidence that 
in the vicinity of the critical size and for eigenvalues close to -1, the
eigenvalues behave in a universal manner controlled by two critical exponents of
$N$, taking the values 1/2 and 3/4. The universality class of this transition is
believed to be that of a random multiplicative ensemble of unitary matrices.
The transition was discovered by Durhuus and Olesen~\cite{duol} when they solved the
Migdal-Makeenko~\cite{makeenko} loop equations in two dimensional planar QCD. We refer 
to it therefore as the DO transition and to the universality class as the
DO class. The multiplicative random 
matrix ensemble~\cite{janik} can be axiomatized in the
language of noncommutative probability~\cite{voicu}. In some sense it provides
a generalization of the familiar law of large numbers, 
which is associated to
the abelian case. The essence of the difference is not in that the 
law of large numbers is additive, 
while the new case is multiplicative; 
rather, the essential features 
making a difference are that 
one case is commutative and the other not. 
Various recent insights into the DO transition
~\cite{olesena, olesenb, blaizot} indicate 
that an even deeper understanding of the transition
might emerge. 

In this work we present numerical results 
indicating that the PCM model
has a similar transition, this time in the two point correlator matrix: when it
is dilated it undergoes a DO transition. Thus, the analogy between the PCM
and gauge theory is upheld and the universal character of the DO transition
as a marker for the transition scale separating weakly from strongly interacting
dynamics in theories based on group manifolds is expanded. 

\section{Basic facts about the PCM.}

The basic degrees of freedom are $g(x)\in SU(N)$ where $x\in R^2$ and the action
is:
\be
S=\frac{N}{T}\int d^2 x Tr \partial_\mu g(x) \partial_\mu g^\dagger (x)
\ee
The large $N$ limit is taken in the 't Hooft prescription, by keeping the
coupling $T$ fixed, which makes $S$ of naive order $N^2$. 

There is a global symmetry group $SU(N)_L \times SU(N)_R$ acting on $g(x)$ by
$g(x)\rightarrow L g(x) R^\dagger$ where $L,R\in SU(N)$. If we eliminate
one of the factors by a translation breaking ``gauge choice'' $g(0)=1$,
we are left with a global symmetry given by a single $SU(N)$ acting on $g(x)$
by conjugation and leaving $g(0)=1$ unchanged. This single $SU(N)$ is
the ``diagonal subgroup'' of the global symmetry group. 

The model is asymptotically free and has a nonzero massgap. Using Bethe ansatz
methods it was found that there are $N-1$ particle states with masses given
by
\be
M_r=M\frac{\sin(\frac{r\pi}{N})}{\sin(\frac{\pi}{N})},~~~1\le r\le N-1
\label{masses}
\ee
Under the diagonal symmetry group, the states corresponding to the $r$-th
mass are a multiplet transforming as an $r$ component antisymmetric tensor. 
This has been verified numerically for $N=6$ in~\cite{numspec}. 
In consequence it is believed that for any $N$ and $1\le r \le N-1$
we would have
\be
\langle \chi_r (g(0) g^\dagger (x) ) \rangle \sim C_r {N\choose r} e^{-M_r |x|}
\ee
for $|x|M >> 1$. Here, $\chi_r$ takes 
the trace in the $r$-antisymmetric representation,
normalized by $\chi_r(1)={N \choose r}$.

The PCM can be formulated in terms of the 
currents $J_\mu(x)=g^\dagger(x)\partial_\mu g(x)$. Note that the
object $g(0)g^\dagger (x)$ is determined by the $J_\mu$ via a path
ordered exponent. In particular, also the action can be expressed in
terms of the $J_\mu(x)$. In addition, the $J_\mu(x)$ obey local
constraints because if viewed as gauge fields they are pure gauge. 

\section{The average characteristic polynomial and the DO transition.}

It was found that a convenient observable to use in order to locate the
DO transition and identify the associated critical $N$-exponents is provided
by the average characteristic polynomial of the unitary matrix undergoing the
transition. This matrix is the open Wilson loop in the gauge case while in the PCM model it is $g(0)g^\dagger (x)$. 

For gauge theory in two dimensions, for a 
non-self-intersecting loop, the unitary matrix can be written as the product
of many such open Wilson loop matrices round small loops tessellating the
spanning area of the loop. For a gauge theory in higher dimensions we 
choose an infinite two dimensional surface extending to infinity and containing
the loop in question and repeat the argument for gauge fields tangent to
the surface. 
The DO transition occurs because for large loops
the correlations between well separated tessellating loops
die out and the factors become effectively independent.
In two dimensions any two distinct tessellating loops are uncorrelated.
Note that the ``number'' of factors in the product is proportional to the
area enclosed by the loop; this product is not to 
be identified with the linearly ordered
product associated with the path ordered exponential. 

For the PCM, we split the segment connecting
$0$ to $x$ into small intervals, with $J_\mu(y)$ determining a multiplicative
contribution for each element. Again, for large $|x|$, the correlations
between well separated small intervals dies out, and a DO transition might
occur. This time however, the ``number'' of factors is proportional to
the linear distance $|x|$. Again, this random product is not to
be identified with the product between the two matrices in the
two-point correlation function. And, yet again, if we descend 
in dimensions (here to one dimension) any two distinct 
intervals are uncorrelated and the DO transition can be established by an exact analytical solution.

In either case, gauge or PCM, 
if $W$ is the unitary matrix
given by the product of independent factors, the
average characteristic polynomial is defined in the same manner:
\be
Q_N(z,\lambda)\equiv \langle \det (z - W) \rangle =\sum_{r=0}^{N} 
z^{N-r} (-1)^r \langle \chi_r (W)\rangle
\ee
Here $\lambda$ is a continuous parameter which has been kept finite
by tuning the distribution of the individual factors in the product
to almost always produce something finite after multiplying the many
factors. In a sense, $\lambda$ would measure the 
area in the gauge case and the
distance in the PCM case. In general, one can view $\lambda$ as a 
parameter quantifying the amount of dilation of the object relative to
a fixed standard. 

In the simple case of two dimensional 
gauge theory one can take $\lambda$ to be
literally the area, made dimensionless with 
the aid of the coupling constant,
and the DO transition can be seen when one uses
\be
\langle \chi_r (W)\rangle = {N\choose r} e^{-\lambda \frac{r(N-r)}{2N}}
\ee

For the PCM, we expect, 
\be
\langle \chi_r (W)\rangle \sim C_r {N\choose r} e^{-M|x| \frac{\sin(\frac{r\pi}{N})}{\sin(\frac{\pi}{N})} }
\ee
We observe that the exponents in the gauge and PCM case are similar, 
with $M|x|$ playing the role of $\lambda/2$. More precisely, for 
$r=O(1)$ or $N-r=O(1)$ and large $N$ they match and for $r\sim N$
they do no differ that much. Thus, if the coefficients $C_r$ vary less than
the other factors, the singularity structure at a critical value of $\lambda$
in the infinite $N$ limit would be the same. 
We conclude that the 
PCM model could also exhibit the DO transition. It would be nice to confirm
this by exact analytical methods, in particular since the numerical
work we shall present indicates that the DO transition does 
occur in the PCM. The commonality between the gauge and PCM case exponents is
related to a topic known as ``Casimir scaling''. 

\section{Regularization and renormalization.}

Until now we ignored renormalization. In both the gauge and PCM case
one needs to renormalize and there is no guarantee that after that it
makes sense to view $W$ as a fluctuating unitary matrix of unit 
determinant. Formally, we can define a renormalized version of the 
$W$ matrix in each case which does provide for such a view. The gauge
case was analyzed in other work, so from now on we restrict ourselves
to the PCM case. 

The extra regularization we put in is of the operator. The action
is regularized with appropriate counter terms in some 
conventional manner. The
net (formal) result is that we assume that we have a probability
distribution generating ``bare'' configurations $g_0 (x)$.

Let $\tau\ge 0$ be a parameter of dimension length squared which means that 
$\tau M^2$ is a finite unitless parameter.  We extend $R^2$ (space time) by an $R^+$ coordinated by $\tau$ and define a
$g(x,\tau)$ determined by $g_0 (x)$ by:
\be
g^\dagger (x,\tau)\frac{\partial g(x,\tau)}{\partial \tau} =\frac{1}{2}[\partial^2 g (x,\tau)-
\partial^2 g^\dagger(x,\tau) ]-\frac{1}{2N} Tr[\partial^2 g (x,\tau)-
\partial^2 g^\dagger (x,\tau) ]
\ee 
with the initial condition
\be
g(x,0)=g_0(x)
\ee
The evolution in $\tau$ ensures that $g(x,\tau)\in SU(N)$ if $g_0(x)\in SU(N)$. 
For $g(x,\tau)=e^{iB(x,\tau)}$ with $ Tr B(x,\tau)=0$ and 
$||B||<<1$ the equation becomes
\be
\frac{\partial B (x,\tau)}{\partial \tau} \approx \partial^2 B(x,\tau)
\ee
This shows that for $\tau >0$ high momentum modes will get suppressed. $g(x,\tau)$ depends nonlocally on $g_0(y)$; the
amount of nonlocality is characterized by the unitless 
finite quantity $\tau M^2$, which is parametrically small 
relative to one, but nonzero. 
As a result, the correlation function 
$\langle g(x,\tau) g^\dagger (y,\tau)\rangle$ will stay finite and nontrivial
in the continuum limit so long as $\tau >0$ 
without needing the extraction
of wave function renormalization factors. 

Hence, the $SU(N)$ matrix $W$ we 
shall compute the average characteristic polynomial of 
will depend on an extra parameter $\tau >0$. The critical
scale $|x-y|_c$ will have some weak dependence on $\tau$. The point is that
so long as $\tau M^2$ is parametrically small relative to unity, there will
be {\sl some} finite $|x-y|_c$ 
where the transition occurs, but the exact value of that $|x-y|_c$ 
is renormalization dependent. The relevant fact is that the DO transition
occurs somewhere and there it marks a sharp boundary
separating strong and weak couplings as defined from this particular
two point function. 

\section{Lattice formulation.}

The continuum formulation ended up with three continuum coordinates. We now discretize each. The two dimensional space-time becomes a square 
lattice with unit lattice spacing, whose sites are again denoted by $x$,
with $x\in Z^2$. $\pm\mu$ denote unit vectors
in the positive/negative $\mu=1,2$ direction.
The $\tau$ direction becomes a unit spaced line
labeled by a coordinate $t=0,1,2,...$. The variables that enter the action
are $g(x)\in SU(N)$ and the $\tau$-dependence of the smeared variables
is denoted by $g_t(x)$, with $g_0 (x) = g(x)$. 
The system is made finite 
by taking the $x$-space to be a torus lattice of equal side lengths, $L$.
The extent in the $\tau$ direction is $n$. Only the space-time direction is
stored in the computer, while the $t$ dependence is iteratively computed.
The $g(x)$ obey periodic boundary conditions and $g_t (x)$ for $t=1,2,...,n$
is determined by $g(x)$ in a manner preserving this.  

The lattice action is:
\be
S_L = -2 N b \sum_{x,\mu} \Re Tr [ g(x) g^\dagger(x+\mu)]
\ee
Averages with respect to
\be
{\cal N} \prod_x [dg(x)] e^{-S_L}
\ee
are denoted by $\langle ... \rangle$ where ${\cal N}$ is a normalization
constant ensuring $\langle 1 \rangle =1$. Thus, the two point function
is
\be
G(x)=\frac{1}{N} \langle Tr [ g(0)g^\dagger(x) ]\rangle
\ee

When $N$ is taken to $\infty$ with $b$ fixed (this is the 't Hooft large $N$ limit)
$G(x)$ remains a finite nontrivial function of $x$. In particular, $\xi_G$,
defined below, measures the correlation in lattice units. 
\be
\xi_G^2 =\frac{1}{4} \frac{\sum_x x^2 G(x)}{\sum_x G(x)}
\ee
$\xi_G$ is independent of the wave function renormalization constant.
It sets the scale for the system in units of the lattice spacing. 
The continuum limit is obtained when $b$ is taken to infinity, 
which causes $\xi_G$ to diverge. 

It has been found that~\cite{numspec} (a number further 
improved in subsequent work) that in the continuum one has
\be
M\xi_G =0.991(1)
\ee
Here, $M$ is the lattice equivalent of the continuum mass $M=M_1$ from Equation~(\ref{masses}). 
We shall use $\xi_G$ as the scale setting quantity. 

The extrapolation to infinite $b$ is greatly helped by using
$E$, the link energy, as an expansion parameter:
\be
E=1-\frac{1}{N}\Re\langle Tr [g(0) g^\dagger({\hat 1} )] \rangle
\ee
As $b\to \infty$, $E\to 0$. This scheme is equivalent to what is also
known as ``mean field improvement''. In the range we shall be working and setting $N$ to infinity, 
the following formula translates $E$ into $b$ to high enough accuracy:
\be
E=\frac{1}{8b}+\frac{1}{256 b^2} +\frac{0.000545}{b^3}- \frac{0.00095}{b^4}+ \frac{0.00043}{b^5}
\ee

The following formula is found to work within a few percent
in the range we shall be working in, which is $11\le \xi_G\le 20$. Again, we use the $N=\infty$ expression, 
but know that $\frac{1}{N^2}$ corrections are small.
\be
\xi_G = 0.991~ \left [ 
\frac{e^{\frac{2-\pi}{4}} }{16\pi} \right ] ~\sqrt{E} ~\exp\left (\frac{\pi}{E}\right )
\ee
Using this formula as setting the scale 
for all dimensionful physical quantities, the approach to continuum can be tested to a level below one percent. 

The smearing operation is discretized as follows:
We start from a
configuration $g(x)\equiv g_0 (x)$ and evolve with a fixed smearing parameter
$0<f<<1$ to a configuration $g_t(x)$. One smearing step takes us from $g_t(x)$ to
$g_{t+1}(x)$. We first define $Z_{t+1} (x)$ by:
\be
Z_{t+1} (x)=\sum_{\pm\mu} [g^\dagger_t (x) g_t (x+\mu)-1 ]
\ee
Next we construct antihermitian traceless $SU(N)$ matrices $A_{t+1} (x)$ 
\be
A_{t+1}(x)=Z_{t+1}(x)-Z^\dagger_{t+1}(x) -\frac{1}{N} \rm{Tr} (Z_{t+1}(x)-Z^\dagger_{t+1}(x))\equiv-A^\dagger_{t+1}(x)
\ee
and set
\be
L_{t+1}(x)=\exp[{f A_{t+1}(x)}]
\ee
Finally, $g_{t+1}(x)$ is defined in terms of $L_{t+1} (x)$ by:
\be
g_{t+1} (x)= g_t (x) L_{t+1} (x)
\ee

This procedure is iterated until $t=n$ is reached. Each iteration is performed
for all $x$ and only then one goes to the next iteration.

For a given $\xi_G$ (or $b$) the parameter $\tau$ is fixed keeping the
quantity $\tau/\xi_G^2$ unchanged; here, $\tau=n f$ is in units of lattice spacing. Again, $\tau$ is the lattice analogue of the
continuum parameter $\tau$. 
Within our range of parameters and with our choice for 
$\tau/\xi_G^2$ we found that 
it is enough to set $n=30$ in order to eliminate a visible dependence
on the two factors $f$ and $n$ individually, making
only $f n$ relevant. This ensures that when
extrapolating to the continuum limit by increasing $\xi_G$ we eliminate
sensitivity on the discretization of the smearing process.

The set of matrices $g(x)$ is generated by Monte Carlo for the action $S_L$.
We set $\xi_G$ and obtain the corresponding $b$. We use a combination of 
Metropolis and Over-Relaxation at each site $x$, where we explore the 
full $SU(N)$ group. Thus, the operation count goes as $N^3$ for
generating configurations. We found that 200-250 passes suffice to thermalize the
system starting from $g(x)\equiv 1$. When going from a configuration
with some $\xi_G$ to another with $\xi_G$ increased by 1, 50 passes are
enough for equilibration and approximate statistical independence.

One major difference between gauge theories~\cite{Narayanan:2003fc,largenred} and the PCM in the context of the
large $N$ limit is that in the PCM there is no special suppression of finite
volume effects. From previous numerical work on the PCM we know that
keeping $L/\xi_G > 7$ reduces finite volume effects to below one percent.
As already mentioned, we carry out simulations in the range $11\le\xi_G\le 20$.
Therefore we chose $L=150$, a relatively large box-size. For smearing we chose
\be
\tau=\frac{\xi_G^2}{300}
\ee
This is a small amount of smearing, but some smearing is necessary in the
continuum limit because we need to avoid the short distance singularity 
one would otherwise expect. A singularity is incompatible with
the $W$ matrices being unitary, and hence bounded, in the continuum limit,

For each $g(x)$ we generate we compute matrices
\be
W_\mu(d)= g_n(x)g^\dagger_n(x+d\cdot\mu)
\ee
for positive and negative directions $\mu$ and all distinct locations.
For a fixed $\xi_G$, our objective is to find $d_c$, the distance
at which the eigenvalue distribution of the $W$'s is critical. For $d<d_c$
the eigenvalue distribution is zero at -1, while for $d>d_c$ it is strictly positive 
everywhere on the unit circle. $d_c$ is fractional, found by interpolation
from the discrete $d$ data. The main objective is to show that  the quantity
$\frac{d_c}{\xi_G}$  has a nontrivial finite continuum limit and to
compute that limit. This would establish the transition as a feature of
the continuum PCM.

\section{The main observable~\cite{ourjhep}.}

As mentioned already, our main observable is related to the 
average characteristic polynomial $Q_N(z,\lambda)$. A closely 
related quantity, $F(y)$, is defined by
\be
F(y,d)=\langle\det(e^{y/2}+e^{-y/2} W)\rangle=C_0 +C_1 y +C_2 y^2 +C_3 y^3 +C_4 y^4+\dots
\ee
We have
\be 
F(y)=\det(W) F^*(-y)=F^*(-y)=F(-y)
\ee
The main reason to introduce $F$ is that it is even in $y$ on account of
$CP$ invariance. This sets $C_1=C_3=...=0$. 

To detect the location of the transition at infinite $N$ we introduce
a derived quantity, $\Omega(d)$, 
which is sensitive to the behavior of $F$ at $y=0$. $y=0$ 
corresponds to $z=-1$ of $Q_N(z,\lambda)$, which is the point at which the
spectrum of the unitary matrix just closes a gap. 
\be
\Omega (d) =\frac{C_0 C_4}{C_2^2}
\ee
$\Omega$ is insensitive to overall rescalings of $F$ and $y$ independently.
In the DO universality class $\Omega$ will jump from being constant at $\frac{1}{6}$ to being constant at $\frac{1}{2}$ at infinite $N$.
This jump will occur at $d=d_c$. The jump in $\Omega (d)$ can be smoothed
out if the critical regime $d\sim d_c$ is dilated as follows:
\be
d-d_c = \frac{\eta}{\sqrt{N}}
\ee
If we take the correlated $N\to\infty$, $d\to d_c$ limit, keeping $\eta$
constant and of order unity, we obtain a smooth function of $\eta$. At $\eta=0$,
the value of $\Omega$ is a universal number:
\be
\Omega_c= \frac{\Gamma^4\left(\frac{1}{4}\right )}{48\pi^2} \approx 0.365
\ee
We now define a finite $N$ approximation to $d_c$, which we denote $d_c(N)$, by
\be
\Omega(d_c(N), N) =\Omega_c
\ee
$d_c$ without an $N$ dependence is the limit $d_c(N=\infty)$. One expects:
\be
d_c (N) = d_c+\frac{\Delta}{N} + O \left(\frac{1}{N^{3/2}}\right)
\ee

In our simulations we obtained numerical values for $\Omega(d,N)$ for
seven values of $d$ centered at $0.65 \xi_G$ and spaced by 1. Using cubic spline
interpolation we found $d_c(N)$ and also an approximation to the derivative
of $\Omega(d,N)$ with respect to $d$ at 
the point $d=d_c(N)$ where $\Omega=\Omega_c$.
We found that this method produces numbers of a relative accuracy of fractions
of a percent for $d_c(N)$ and a few percent for the derivative. 
Every input set of 7 $\Omega$-values for fixed $\xi_G$ and $d$ was 
obtained by averaging the coefficients $C_k$ over all possible $g$-pairs
and subsequently averaging the $\Omega$-s obtained from
these average values of $C_k$ for fixed configurations of $g$
over about 50 different and statistically independent $g$-configurations. The errors
in the $\Omega$-s were obtained by jackknife with single elimination. 

Henceforth, we shall redefine $\Omega$ to be $\Omega -\Omega_c$.
Hence $d_c (N) $ will be associated with the vanishing of the new $\Omega$.

\section{Results.}

\FIGURE{
\epsfig{file=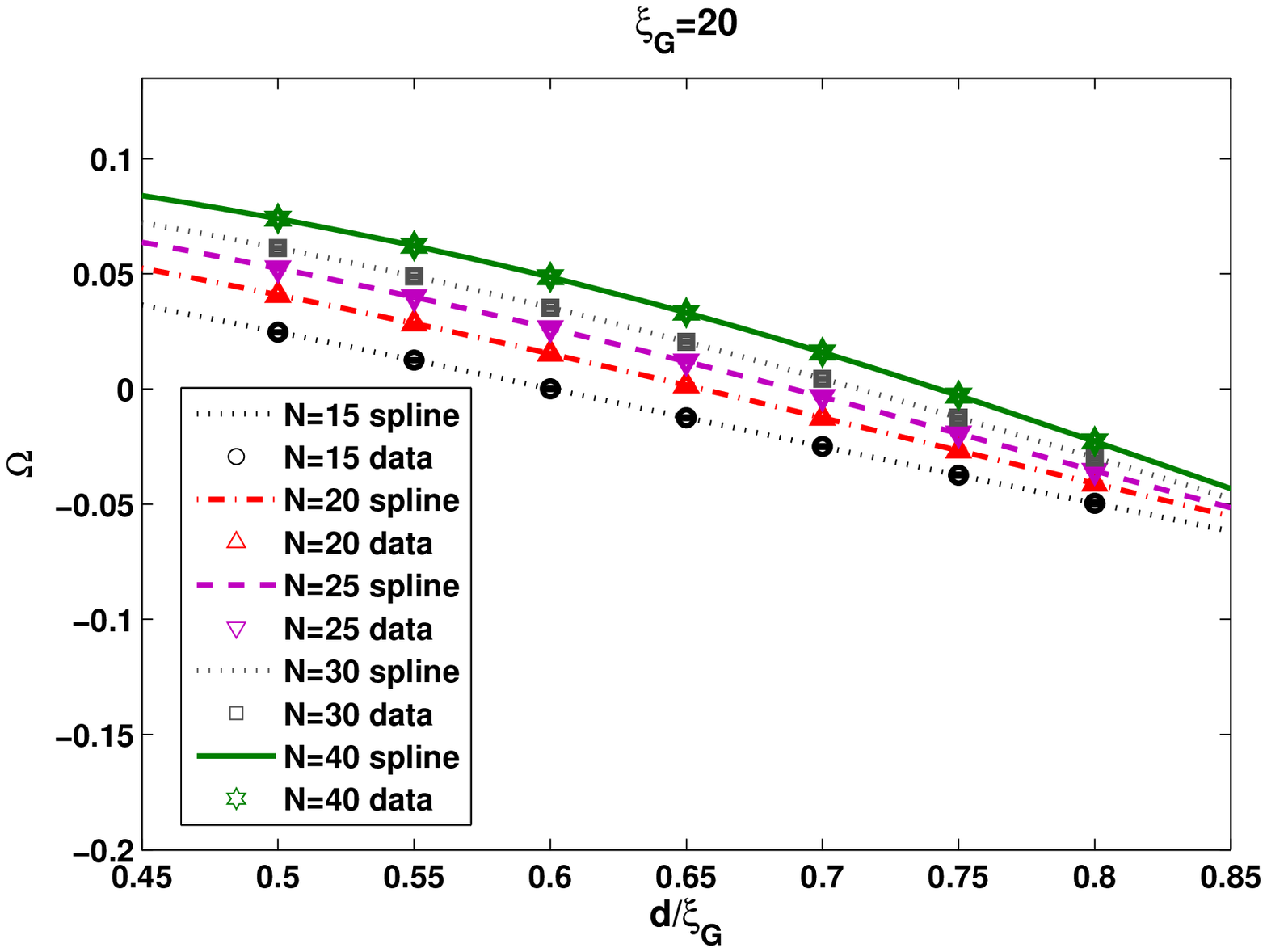, height=3.5in }
\caption{Plot of the subtracted $\Omega$ as a function of $\frac{d}{\xi_G}$ for 
$N=15,20,25,30,40$ and for the 
finest lattice, with $\xi_G=20$. The data point sizes are much larger than the statistical 
error bars which might be visible after
significant magnification. }
\label{cubspline}}

\FIGURE{
\epsfig{file=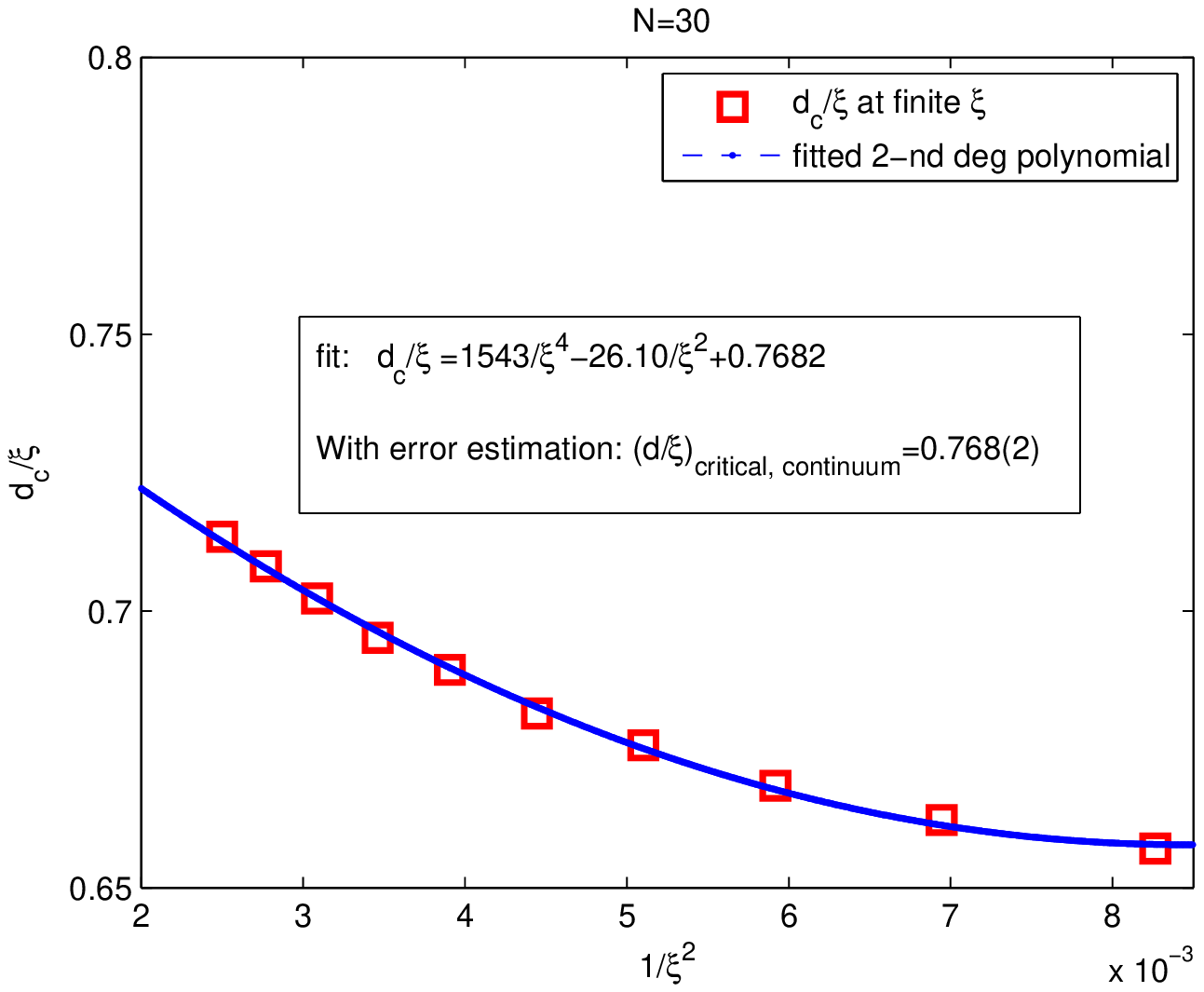, height=3.5in }
\caption{Extrapolation to continuum of $\frac{d_c(N)}{\xi_G}$ for N=30.  }
\label{extrp}}

\FIGURE{
\epsfig{file=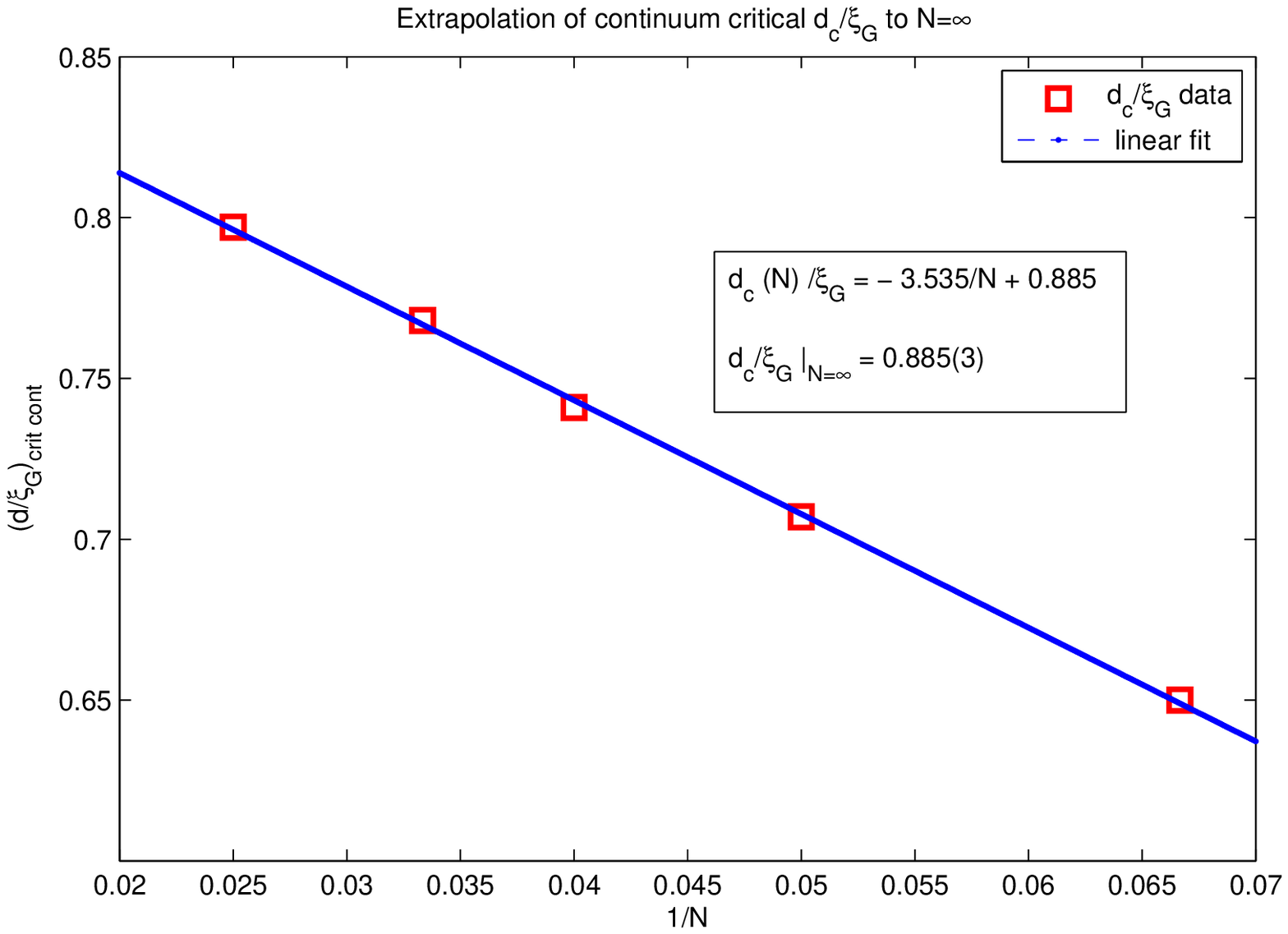, height=3.5in }
\caption{Extrapolation of $\frac{d_c(N)}{\xi_G}\left|_{\rm continuum}\right.$  to
infinite $N$. }\label{critd}}

\FIGURE{
\epsfig{file=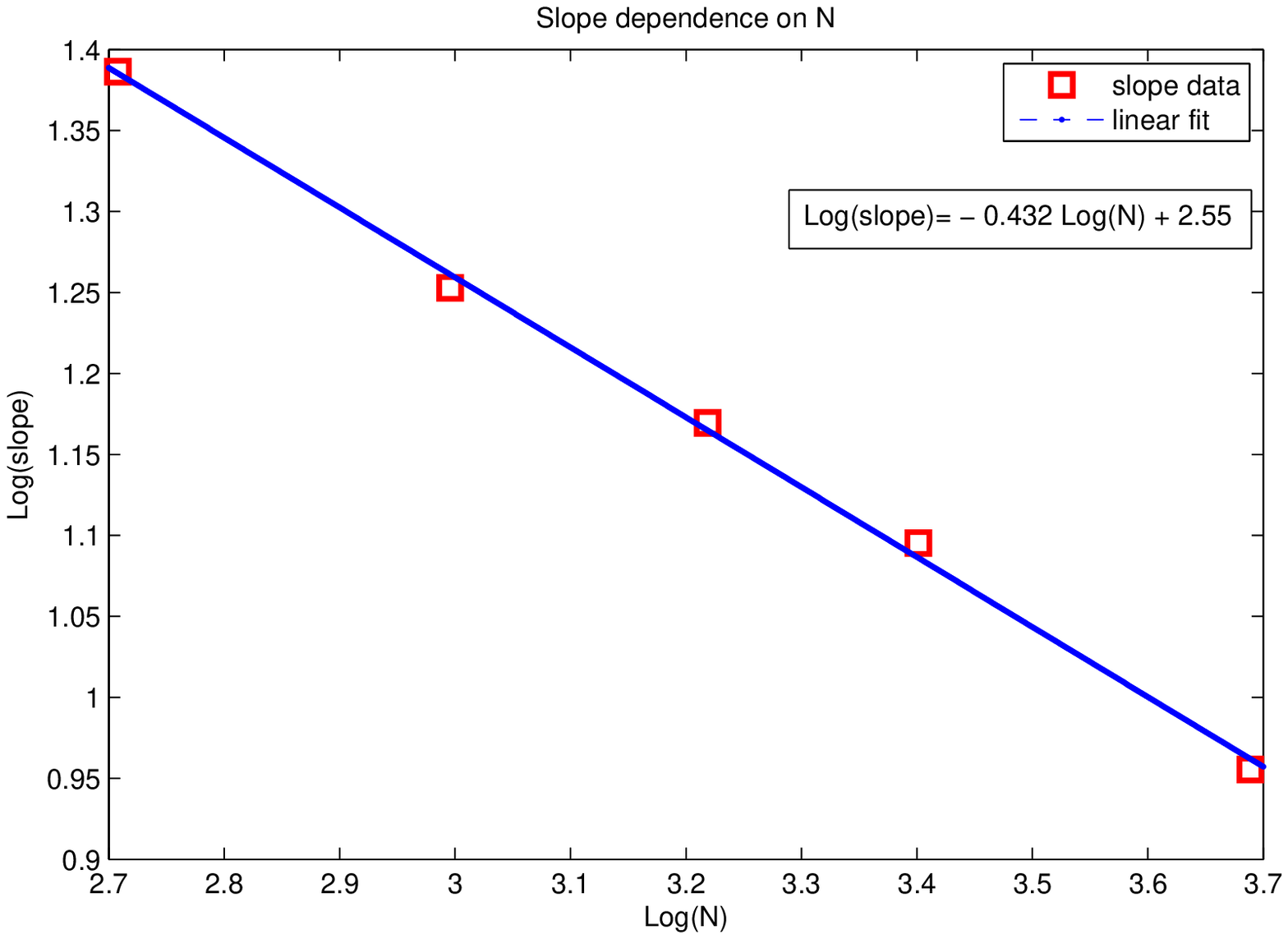, height=3.5in }
\caption{Testing for the critical exponent 1/2 in the slope
variable. }\label{slope}}

For each $\xi_G$ in the range of 11 to 20 we end up with a $d_c(N)$ and a slope
of $\Omega$ at $\Omega=0$. To check for a continuum limit we look at the ratios
$\frac{d_c(N)}{\xi_G}$. For each value of $N$ we extrapolate to the continuum
by fitting the data for $\frac{d_c(N)}{\xi_G}$ to a second order polynomial in
$\frac{1}{\xi_G^2}$. This 
extrapolation is seen to be quite
smooth and the second order polynomial provides a good fit. The slopes on the
other hand have too low an accuracy to permit an extrapolation to the continuum
limit; within their errors of a few percent they turn out to be $\xi_G$ independent.
We took for the slopes the value on the finest lattice we have, at $\xi_G=20$ and
carried out no continuum extrapolation.  

Finally, we are ready to look at the $N$ dependence. For $d_c(N)$ we expect
a convergence to an infinite $N$ linear in $\frac{1}{N}$. The slope
of $d(N)$ with respect to $\Omega$ at $\Omega=0$ should go to zero as $\frac{1}{\sqrt{N}}$. Sample figures are below.

Figure~\ref{cubspline} shows how the data for $\Omega$ is interpolated to get
the critical size for a particular lattice spacing (the example is for $\xi_G=20$)
at various values of $N$. 
Note that $\Omega$ gets steeper as $N$ increases. 
The data points markers are 
larger than the statistical jackknife errors.

Figure~\ref{extrp} shows the extrapolation to the continuum for $N=30$. 
The discrete value are plotted as squares to stand out, but the sizes 
of the squares do not indicate the errors. The error on the extrapolated
value was computed by generating artificial noise on the input data with
Gaussian errors of size determined by the jackknife estimates of the
configuration averages. 40 such fake sets were next interpolated by 
cubic spline and then extrapolated to continuum to generate the error 
on the extrapolated value shown in the figure.

In Table~\ref{tab} we summarize our findings. The definition of the entry
``slope'' is:
\be
\left [ \frac{d\left ( d_c(N)/\xi_G \right ) }{d\Omega} \right ]_{\Omega=0} 
\ee

\TABLE{
\begin{tabular}{ccc}
$N$ & $\left ( \frac{d_c}{\xi_G}\right )_{continuum}$ & slope  \cr
\hline
15 &  0.650(3) & -4.00\\
20 &  0.707(2) & -3.58\\
25 &  0.741(2) & -3.22\\
30 &  0.768(2) & -2.99\\
40 &  0.797(2) & -2.60\\
\hline
\end{tabular}
\caption{\label{tab} Input to the extrapolation of $N$ to $\infty$}}

The final objective is to extrapolate to infinite $N$. 
For the critical size this is shown in Figure~\ref{critd}.
We see that a simple linear extrapolation in $\frac{1}{N}$ works well and 
obtain in the infinite $N$ limit
\be
\frac{d_c}{\xi_G}\left |_{continuum; ~\tau=\frac{\xi_G^2}{300}} = 0.885(3)\right.
\ee

Even if we accept that the $d_c(N)$ that we defined
has a limit as described, we have presented so far little hard
evidence that the transition is of the DO type. To see this, we look at the
slope and check whether numerically it is plausible that it exhibits a dependence
on $\sqrt{N}$ as the DO universality class would predict. 
In Figure~\ref{slope} we show a plot of $\log|slope|$ versus $\log(N)$.
One sees a slope that is different from the 
expected -0.5, but we have not included any
further subleading corrections in $\frac{1}{N}$. 
There is little doubt that a slope
of -1 is excluded by the data and the data is consistent with a slope of -0.5 in the asymptotic large $N$ regime. If we drop the lowest $N$ value from the fit the slope increases somewhat, but does not reach 0.5.
The slopes for finite $N$ had too little accuracy to allow
a continuum extrapolation and therefore we used only
the slopes at $\xi_G=20$, our largest correlation length. 
There is no reasonable way to get a real error on our slope
result, so no errors were quoted. Still, our result supports
the expectation that the critical exponent associated with $d-d_c$ is 1/2.

Nothing we have done so far can be a substitute to a simple direct comparison of
the DO eigenvalue distribution to the data. In Figure~\ref{doa} and Figure~\ref{dob} we show
for N=40 plots of the eigenvalue density we obtained 
at $\xi_G=20$ for $d=10,21$ and $d=17$ respectively. 
Superposed on the data in both figures are the DO distributions obtained by
fitting their single free parameter $k$. $k$ gives a measure for
how far we are from the critical point which corresponds to
$k_c=2$. 
Durhuus and Olesen obtained an expression for the 
distribution of the angle $\theta$~\cite{duol}. The eigenvalues of
$W$ are $e^{\imath\theta_j}$, with $\sum_j \theta_j =
0(mod)2\pi$ and $p(\theta)d\theta$ is the
probability that any $\theta_j$ satisfy $|\theta_j -\theta|<d\theta/2$. $p(\theta)$ is defined from a complex
function $h(k,\theta)$ which depends parametrically on $k$:
\be
p(\theta) = -\frac{1}{2\pi k} {\rm Im\ } h(k,\theta).
\ee

\FIGURE{
\epsfig{file=doa.eps, height=4.0in }
\caption{Examples of the eigenvalue distribution for a small ($d\sim 0.5\xi_G$) and a large distance ($d \sim \xi_G$).}\label{doa}}

The complex valued function $h(k,\theta)$ is determined by a
nonlinear equation. One first introduces two $\theta$ 
dependent real variables, $x$ and $y$, and the expressions
$f(y)=\frac{y}{\sinh(y)},~g(y)=y\coth (y)$.  
\be
x = - f(y) \sin (x+\theta), ~~
x+\theta = {\rm Re\ }h(k,\theta),~~
y = {\rm Im\ }h(k,\theta)
\ee
These equations change variables from $h,\theta$ to $x,y,\theta$. The dependence on $k$ comes in through the
main equation
\be
g(y) - k = f(y) \cos (x+\theta)
\ee
$k$ determines the support of
$p(\theta)$, which just reaches $\theta=\pm\pi$ at $k=k_c$. 
\be
|\theta| \le \cases {\pi & if $k>2$ \cr
\cos^{-1}(1-k) +\sqrt{2k-k^2} & if $k < 2$\cr}
\ee
 
\FIGURE{
\epsfig{file=dob.eps, height=4.0in }
\caption{An example of an almost critical eigenvalue distribution ($d\sim 0.85\xi_G$).}\label{dob}}

In Figure~\ref{doa} we see two extreme cases, one corresponding
to a gapped eigenvalue distribution for a distance of the order
$d=0.5\xi_G$ and the other for an eigenvalue distribution covering the entire unit circle, for a distance $d=1.05 \xi_G$. 
The solid lines are the exact DO distribution, at a $k$ that was
determined by the best least squares fit to the empirical histogram.
Taking data for a larger set of distances then usual, we can
ask which distance provides a $k$ closest to critical. It turns
out that this happens at $d=17$, where $k=1.98$. Thus, this
simple method would have given us an estimate for the infinite
$N$ critical ratio $\frac{d_c}{\xi_G}$ 
of about $ 0.85$. 
Our result for our definition for a finite $N$ critical
distance $d_c(N)$ for $N=40$ gave $\frac{d_c}{\xi_G}\sim 0.80$
at $\xi_G=20$ while the continuum infinite $N$ number was 
0.885. This is all consistent. Figure~\ref{dob} shows this close
to critical distribution. 

One may speculate that the DO distribution matches the
continuum infinite $N$ distribution exactly in the critical case, but this is very likely not correct. Universality
holds only for $\theta$ close to $\pm\pi$. If one looks at the
integrated 
square distance between the DO distribution and the empirical
one as a function of $N$ there is no evidence that this
distance extrapolates to zero.

\section{Summary.}

There is little doubt that with the introduction of the smearing parameter $\tau$
the PCM undergoes an infinite $N$ phase transition of the DO type, in a manner
analogous to smeared Wilson loops in two, three and four dimensions. 

The PCM also offers the hope to establish this transition by analytic means.
If this could be done, the role of the parameter $\tau$ 
could be further elucidated. We look at it as an extra renormalization of the
two point matrix needed to reconcile its short distance behavior with it
being a fluctuating unitary (and hence bounded) matrix. 

At infinite $N$ the PCM model has a a trivial S-matrix for its set of
bound states. Nevertheless, it is not a free field theory in terms of the
variables $g(x)$ as evidenced by the short distance singularity of the two 
point function. The introduction of $\tau$ eliminates this divergence but
since $\tau$ can be made very small, the short distance behavior of the
two point function can be seen for $\sqrt{\tau} << d << \xi_G$. The DO transition
occurs at $d \sim \xi_G$ and separates the two regimes, one of noninteracting
bound states and the other of Gaussian fluctuations of the fundamental
variable $g(x)$.

\acknowledgments

R. N. acknowledge partial support by the NSF under grant number
PHY-055375.  H. N. acknowledges 
partial support by the DOE under grant
number DE-FG02-01ER41165 at Rutgers University. H. N. also 
notes with regret that his research has for a long time been 
deliberately obstructed by his high energy colleagues at Rutgers.

\end{document}